\documentclass[10pt,conference]{IEEEtran}
\IEEEoverridecommandlockouts
\usepackage{cite}
\usepackage{amsmath,amssymb,amsfonts}
\usepackage{graphicx}
\usepackage{textcomp}
\usepackage{xcolor}
\usepackage{booktabs}       
\usepackage[utf8]{inputenc} 
\usepackage[T1]{fontenc}    
\usepackage{hyperref}       
\usepackage{url}            
\usepackage{booktabs}       
\usepackage{amsfonts}       
\usepackage{nicefrac}       
\usepackage{microtype}      
\usepackage{neuralnetwork}
\usepackage{subcaption}
\usepackage{comment}
\usepackage{url}
\usepackage{algorithm}
\usepackage{algpseudocode}
\usepackage{listings}
\usepackage{multirow}
\usepackage{balance}

\def\BibTeX{{\rm B\kern-.05em{\sc i\kern-.025em b}\kern-.08em
    T\kern-.1667em\lower.7ex\hbox{E}\kern-.125emX}}

\newcommand{\AF}{Action Filter}

\newcommand{\IC}{InspectorClone}

\newcommand{\tone}{Type I}
\newcommand{\ttwo}{Type II}
\newcommand{\tthree}{Type III}
\newcommand{\tfour}{Type IV}

\newcommand{\thickhline}{\noalign{\hrule height 1.0pt}}
\newcommand{\SA}{semi-automated}

\makeatletter

\def\therule{\makebox[\algorithmicindent][l]{\hspace*{.5em}\vrule height .75\baselineskip depth .25\baselineskip}}%

\newtoks\therules
\therules={}
\def\appendto#1#2{\expandafter#1\expandafter{\the#1#2}}
\def\gobblefirst#1{
	#1\expandafter\expandafter\expandafter{\expandafter\@gobble\the#1}}%
\def\LState{\State\unskip\the\therules}
\def\pushindent{\appendto\therules\therule}%
\def\popindent{\gobblefirst\therules}%
\def\printindent{\unskip\the\therules}%
\def\printandpush{\printindent\pushindent}%
\def\popandprint{\popindent\printindent}%

\algdef{SE}[WHILE]{While}{EndWhile}[1]
{\printandpush\algorithmicwhile\ #1\ \algorithmicdo}
{\popandprint\algorithmicend\ \algorithmicwhile}%
\algdef{SE}[FOR]{For}{EndFor}[1]
{\printandpush\algorithmicfor\ #1\ \algorithmicdo}
{\popandprint\algorithmicend\ \algorithmicfor}%
\algdef{S}[FOR]{ForAll}[1]
{\printindent\algorithmicforall\ #1\ \algorithmicdo}%
\algdef{SE}[LOOP]{Loop}{EndLoop}
{\printandpush\algorithmicloop}
{\popandprint\algorithmicend\ \algorithmicloop}%
\algdef{SE}[REPEAT]{Repeat}{Until}
{\printandpush\algorithmicrepeat}[1]
{\popandprint\algorithmicuntil\ #1}%
\algdef{SE}[IF]{If}{EndIf}[1]
{\printandpush\algorithmicif\ #1\ \algorithmicthen}
{\popandprint\algorithmicend\ \algorithmicif}%
\algdef{C}[IF]{IF}{ElsIf}[1]
{\popandprint\pushindent\algorithmicelse\ \algorithmicif\ #1\ \algorithmicthen}%
\algdef{Ce}[ELSE]{IF}{Else}{EndIf}
{\popandprint\pushindent\algorithmicelse}%
\algdef{SE}[PROCEDURE]{Procedure}{EndProcedure}[2]
{\printandpush\algorithmicprocedure\ \textproc{#1}\ifthenelse{\equal{#2}{}}{}{(#2)}}%
{\popandprint\algorithmicend\ \algorithmicprocedure}%
\algdef{SE}[FUNCTION]{Function}{EndFunction}[2]
{\printandpush\algorithmicfunction\ \textproc{#1}\ifthenelse{\equal{#2}{}}{}{(#2)}}%
{\popandprint\algorithmicend\ \algorithmicfunction}%
\makeatother

\begin{document}

\title{Towards Automating Precision Studies of Clone Detectors}



\author{

	Vaibhav Saini\textsuperscript{\textdagger} ~~~~~~~~~~
	Farima Farmahinifarahani\textsuperscript{\textdagger}~~~~~~~~~~
	Yadong Lu\textsuperscript{\textdagger}\\ 
	Di Yang\textsuperscript{\textdagger}~~~~~~~~
	Pedro Martins\textsuperscript{\textdagger}~~~~~~~~
	Hitesh Sajnani\textsuperscript{\textdaggerdbl}\\
	Pierre Baldi\textsuperscript{\textdagger}~~~~~~~~
    Cristina V. Lopes\textsuperscript{\textdagger}\\
	\IEEEauthorblockA{\textsuperscript{\textdagger} UC Irvine, USA ~~~~~~~~~~ 	\textsuperscript{\textdaggerdbl} Microsoft, USA}\\
	\text{\{vpsaini, farimaf, yadongl1, diy4, pribeiro, pfbaldi, lopes\}@uci.edu} \\

	\text{hitsaj@microsoft.com}
}


\maketitle

\begin{abstract}
Current research in clone detection suffers from
poor ecosystems for evaluating precision of clone detection tools.
Corpora of labeled clones are scarce and incomplete, making
evaluation labor intensive and idiosyncratic, and limiting inter-tool
comparison. Precision-assessment tools are simply lacking.

We present a \SA\ approach to facilitate precision studies
of clone detection tools. The approach merges automatic mechanisms of clone classification with manual validation of clone pairs. We demonstrate that the proposed automatic approach has a very high precision and it significantly reduces the number of clone pairs that need human validation during precision experiments. Moreover, we aggregate the individual effort of multiple teams into a single evolving dataset of labeled clone pairs, creating an important asset for software clone research.
\end{abstract}


\section{Introduction}
Source code clone detection is the task of finding similar software
pieces, according to a certain concept of similarity. These pieces can
be statements, blocks of code, functions, classes, or even complete
source files, and their similarity can be syntactic, semantic or
both. Cloning in software source code is as ubiquitous as software
itself, which gives clone detection tools many applications:
plagiarism and copyrights enforcement~\cite{hummel2010index,roy2009comparison}, detection of
errors/faults/bugs~\cite{roy2009comparison}, code optimization and
refactoring~\cite{weissgerber2006identifying,kawaguchi2009shinobi},
analysis of programmers behaviors~\cite{yamashina2008shinobi} or
program understanding~\cite{roy2009comparison} are some examples.

In a systematic literature review, Rattan et al. found at least 70 clone detection tools and techniques~\cite{RATTAN20131165}. Clone detectors differ substantially in the underlying techniques and
scope of application. In terms of technical approach used, one can find techniques that are learning-based~\cite{white2016deep,oreopreprint}, token-based~\cite{li2006cp,sajnani2016sourcerercc,svajlenko2017fast}, tree-based~\cite{jiang2007deckard,baxter1998clone}, graph-based~\cite{gabel2008scalable}, or text-based~\cite{roy2008nicad}. With respect to the scope, one can find tools that are language-specific and language-agnostic, with varying degrees of specialization. While there are many tools and techniques published to detect clones, not much effort is spent on streamlining the evaluation of these tools and techniques. 

The effectiveness of clone detection tools is usually evaluated in terms of precision and recall. Precision is the percentage of true positives (clone pairs)
within a set of code pieces identified by the tool as clones. Recall
is the percentage of true positives that are retrieved by the tool
within the complete set of known clones. The measurement of precision
and recall, in general, relies on the existence of labeled datasets. 
A good labeled dataset provides realistic
data and credible labels on all the constituents that should be
detected by the analysis tool -- in the case of clone detection, all
clone pairs are labeled as such. A labeled dataset for clones allows
one to measure how many of the clones identified by a certain tool are
indeed clones or not (precision), and how many of the true clone pairs
are detected by the tool (recall). Publicly available labeled datasets, also known as benchmarks, allow direct inter-tool comparisons without the uncertainty that
exists when two tools are compared with different datasets.

In the field of code clone detection, building labeled
datasets is particularly challenging and requires software expertise. Methods can exist inside methods, or they can vary wildly in
size, scope, semantics and nature. In addition, to manually validate all the possible clones would require quadratic comparisons, a combinatorial
problem that becomes infeasible with growing codebases. 

For this reason, most datasets used in code clone studies are either small or synthetically created or are labeled only for a subset
of pairs. The dataset by Bellon \textit{et al.}~\cite{bellon2007comparison}, the dataset by Murakami \textit{et al.}~\cite{murakami2014dataset}, SOCO 2014~\cite{soco} or BigCloneBench~\cite{Svajlenko:2014:TBD:2705615.2706134} have one of the above mentioned limitations.

There has been good progress in measuring recall
systematically of clone detection tools. Based on BigCloneBench
dataset~\cite{Svajlenko:2014:TBD:2705615.2706134},
BigCloneEval~\cite{7816515} estimates recall automatically by
measuring how many of the labeled clone pairs are included in the
output of a clone detector. However, BigCloneEval stops short of
estimating precision because BigCloneBench does not contain labels for
{\em all} possible clone pairs in it. If a clone detector identifies a
clone pair that is not marked as such, only manual inspection can tell
whether the pair is a false positive or a true positive. Since manual inspection is a difficult, labor intensive, and time consuming task, most clone detection approaches estimate their precision by sampling a number of their reported clone pairs, and then manually inspecting the sampled set~\cite{sajnani2016sourcerercc, oreopreprint, Wang:2018:CTB:3180155.3180179}. So, while clone detectors report recall using BigCloneEval, the determination of
their precision is still a subjective and a manual process, leading to difficulties in
comparing with other tools. 
The lack of an established labeled dataset
for precision creates a number of problems~\cite{sajnani2016sourcerercc}: (i) precision estimation can suffer from sampling bias when sample set is small and not representative of the population, (ii) large manual effort is required to conduct precision studies. As each pair in the sample needs to be evaluated, often by multiple judges, the total manual effort required to complete a precision study is substantial, and (iii) the efforts put into the manual inspection of clone pairs are
not reused. Each time authors want to estimate their tool's  precision, they typically start from scratch.

To address these problems we present
\IC, an approach designed to facilitate precision studies for code
clone detectors. \IC\ helps in evaluating precision of clone detectors and, in the process, creates a dataset of well-known
source code clones. \IC\ \textit{automatically} resolves as many clone pairs as possible, identifying a subset of pairs where manual inspection is most needed. Our experiments demonstrate that \IC\ reduces the number of clone pairs that need manual inspection by 40\% on an average.

At the end, the results (automatic and manual) are
aggregated to report the precision of the clone detector. Moreover, the
human judgments are stored to create a manually labeled dataset of clones.
This dataset is beneficial in 
inter-tool comparison, and also, for the exploration of machine learning and
artificial intelligence techniques in clone detection.
InspectorClone can be accessed at~\url{http://www.inspectorclone.org}.

The main contributions of this work are the following: (i) a \SA, high-precision, approach for classification of clones that reduces the manual effort of precision studies significantly, (ii) a publicly available web application that enables clone detection researchers to conduct precision experiments of clone detection tools and techniques, and (iii) an evolving dataset of manually validated clone pairs that is publicly available. With time, as \IC\ is used, the number of humanly validated clone pairs will increase in the dataset.


This work is organized as follows. In Section~\ref{sect: tool}, we
present \IC, the tool implementation of our approach. The automatic mechanisms of clone pair resolution, and related concepts are elaborated in Section~\ref{sect:automatic}, and then it is evaluated in Section~\ref{sect:eval}. Section~\ref{sect:related} presents related work, and threats to
validity are explained in Section~\ref{sect:threat}. Finally, we
present the conclusions and future work in
Section~\ref{sect:conclusion}.

\section{InspectorClone}
\label{sect: tool}
\begin{figure*}[t]
	\centering
	\includegraphics[scale=0.4]{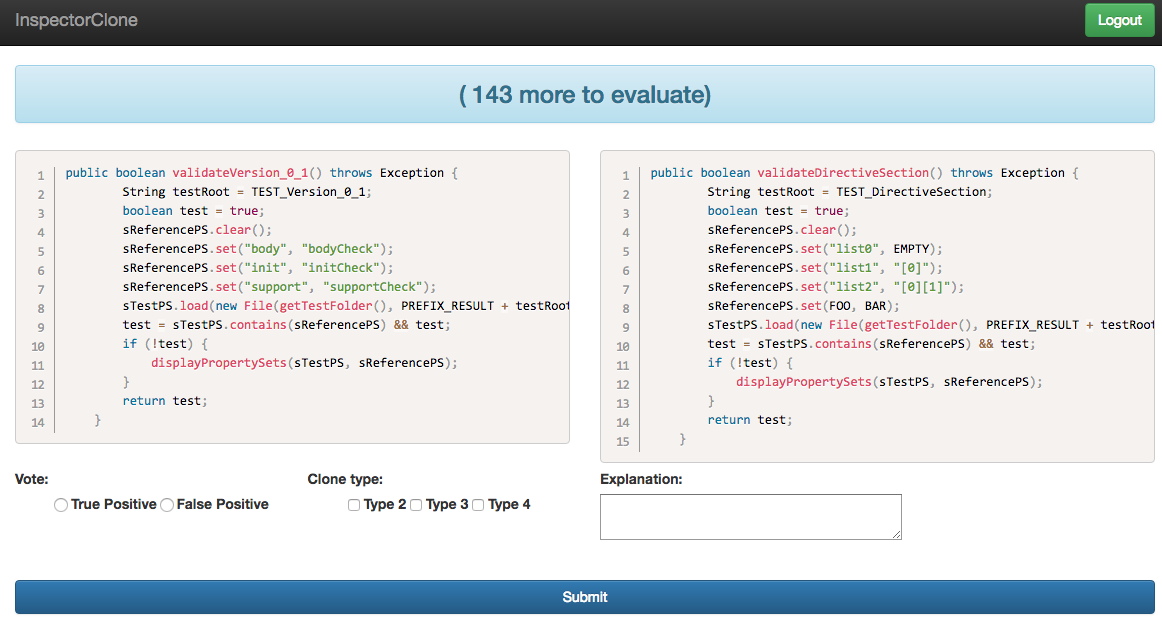}
	\caption{Validation of clone candidates on InspectorClone.}
	\label{fig:evaluate_pairs_screen}
\end{figure*}

Measuring the precision of a clone detector is not a trivial process. To measure the precision, one can choose to manually validate all the clone pairs reported by a clone detector. This process, however, is extremely time consuming and impractical as the number of clone pairs reported by a tool on a standard dataset like BigCloneBench is in millions. A more practical process is to estimate the precision by humanly validating a random and statistically significant sample of clone pairs. This is what researchers do to estimate the precision of clone detectors~\cite{sajnani2016sourcerercc,Svajlenko:2015:ECD:2881297.2881379,Wang:2018:CTB:3180155.3180179,oreopreprint}. In this process, after running a clone detector on a dataset and getting the clone pairs, a random and statistically significant sample set of these clone pairs is assigned to multiple judges for manual
inspection. The judges examine each pair to decide if it is a true clone
and/or what type of clone it is. When all sampled pairs have been
validated by all judges, researchers aggregate the judges'
decisions, usually by taking the majority vote, and report precision. 


The above process, though more practical than humanly validating every clone pair, still takes a non trivial amount of time and effort. Moreover, the effort put into one study cannot be reused in future studies. To address these issues, we present a web-based tool, named \IC, that helps clone researchers in expediting the precision estimation process. \IC\ helps by mimicking this whole process and also by automatically validating a subset of sampled clone pairs, thereby reducing the number of clone pairs shown to human judges. Moreover, by storing human judgments in a centralized database, this tool turns humans' manual effort to a long lasting resource that can be reused in future studies.


\IC\ conducts precision studies on the dataset curated by Svajlenko et al. for facilitating recall studies~\cite{7816515}. Svajlenko et al. curated this dataset using IJaDataset-2.0 to conduct recall study using BigCloneEval. The dataset is available for download on \IC's website. \IC\ does not run the clone detection tool; instead, it expects users to upload the clone pairs reported by their tool to the website. The work flow is as follows. 

A user, John, registers himself and his tool into \IC. After registration, John can download the dataset of source code and run it
on his clone detector. John, then uploads the clone pairs to \IC\ where  \IC\ filters out the methods that are less than 50 tokens, a standard filter used in precision studies~\cite{sajnani2016sourcerercc, Svajlenko:2015:ECD:2881297.2881379}. John now creates an experiment to estimate the precision of his tool. He then invites multiple judges to evaluate the pairs. Once the judges are invited, \IC\ selects a random and statistically significant sample of clone pairs. From this sample \IC\ tries to automatically validate as many pairs as it can. All of the remaining pairs of the sample, which \IC\ did not resolve, are then shown to the judges.

When a judge, Alice, starts an experiment assigned to her, she is
shown a web page as shown in
Figure~\ref{fig:evaluate_pairs_screen}. All unresolved pairs will be
shown to her.  This page is composed by a split
screen with two columns, showing both members of a pair. The
code is syntax highlighted to increase the
readability. Alice must then decide if this pair does indeed represent
a clone or not (if it is a true or a false positive). There are two
optional form elements: one to select the clone type, and another to
leave a comment. 

When all of the unresolved pairs have been validated by all judges,
\IC\ aggregates their decisions by taking the majority vote, and
creates a precision report. In case there are even number of judges,
\IC\ treats a pair as a true positive only when more than 50\% of the
judges vote for it to be a clone pair. \IC\ stores the human judgments
in a centralized database. With time, we expect the number of humanly
judged pairs to increase in this database, thereby creating a valuable
asset for the community. 

\section{Automatic Classification of Clones}
\label{sect:automatic}

As explained in the previous section, to estimate the precision of a clone detection tool, \IC\ needs to validate only a random and statistically significant sample set of the clone pairs reported by the tool. The number of these pairs in such a sample set is small and therefore, \IC\ can use techniques which are very precise without caring much about the scalability aspects of the techniques. 

Also, an automatic approach must be able to resolve pairs with very high precision; otherwise, researchers will fall back to the completely manual process. With this in mind, we designed a \SA\ approach to conduct precision studies using \IC. The automatic mechanism of \IC\ has a very high precision but it compromises on recall as it only resolves those pairs on which it has high confidence. The unresolved pairs are then shown to human judges for manual inspection.
We note that clone detection tools operate at various granularities
like statements, block of code, methods, files, et cetera. Also, clone
detection can be carried out for software written in various languages
like Java, C, C++, and Python among others. In this work, we narrow down
our focus to facilitate precision studies for method level clone
detectors which find clones in software systems written in Java.
%
%
%
%

\subsection{Definitions}
\label{sect:defs}
In this section, we elaborate on the terms and definitions that are pivotal to discussing \IC's mechanisms.

\textbf{Clone Pair}: A pair of code fragments that are similar,
specified by the triple (f1, f2, $\phi$), including the similar code
fragments f1 and f2, and their clone type $\phi$~\cite{Svajlenko:2015:ECD:2881297.2881379}.

\textbf{Clone Types}: Based on the literature~\cite{roy2009comparison}, our work uses the following
four types of source code clones, the first three being similar on the
textual and syntactic level, and the fourth type defining similarity
on the functional, semantic level:

\textbf{\tone}: Identical code fragments, except
	for differences in white-space, layout and comments.
	
\textbf{\ttwo}: Identical code
	fragments, except for differences in identifier names and literal
	values, as well as \tone\ differences.
	
\textbf{\tthree}: Syntactically similar code
	fragments that differ at the statement level. The fragments have
	statements added, modified and/or removed with respect to each
	other, in addition to \tone\ and \ttwo\ clone differences
	
\textbf{\tfour}: Syntactically dissimilar code fragments that implement the same functionality. 
	
The definition to classify clones as \tthree\ does not specify what should be the minimum syntactical similarity between the methods of a clone pair to be classified as \tthree. Also, the lack of consensus in the community of clone researchers about this similarity makes it difficult to separate \tfour\ and \tthree\ clones. To address this issue, the popular clone benchmark, BigCloneBench~\cite{Svajlenko:2014:TBD:2705615.2706134,
	Svajlenko:2015:ECD:2881297.2881379}, has divided the zone between \tthree\ and \tfour\ into four subcategories based on syntactical similarity values: Very Strongly~\tthree\ (VST3) with similarity in
range of [0.9, 1.0), Strongly~\tthree\ (ST3) with similarity being in [0.7,0.9),
Moderately~\tthree\ (MT3) with similarity in range of [0.5, 0.7), and Weakly~\tthree\ (WT3/4) having similarity in the range of [0.0,0.5). More details about these subcategories can be found
elsewhere~\cite{Svajlenko:2014:TBD:2705615.2706134}.

\textbf{Action Token}: Action tokens of a
method are the tokens corresponding to the methods called and class fields
accessed by that method~\cite{oreopreprint}. Additionally, the array accesses made by a method are also special Action tokens namely \textit{ArrayAccess} and \textit{ArrayAccessBinary}, where array access of kind \textit{arr[i]} is an \textit{ArrayAccess} Action token and \textit{arr[i+1]} is an \textit{ArrayAccessBinary} Action token. In the code provided in Listing~\ref{lst:actiontokens}, Action tokens are: \textit{children()}, \textit{hasMoreElements()}, \textit{nextElement()}, \textit{isFiltered()}, \textit{addElement()}, and \textit{elements()}.

\textbf{Action Filter}: A filter which ensures a minimum amount of similarity between the Action tokens of two methods~\cite{oreopreprint}. 

We use overlap-similarity, calculated as $Sim(A_1,A_2)= |A_1\cap A_2|$, to measure the similarity between the
Action tokens of two methods. Here, $A_1$ and $A_2$ are sets of
Action Tokens in methods $M_1$ and $M_2$, respectively. Each element in
these sets is defined as $<t,freq>$, where $t$ is the Action Token and
$freq$ is the number of appearances of this token in the method. $M_1$ and $M_2$ satisfy the Action filter if $ \frac{Sim(A_1,A_2)}{ max(|A_1|,|A_2|)}$$\geq$$\theta$, where $\theta$ is Action filter threshold such that $0 \leq \theta \leq 1$.

\begin{lstlisting} [label={lst:actiontokens}, float,floatplacement=H,caption=Example: Action Tokens] 
public Enumeration children() {
	Enumeration allChildren = super.children();
	Vector filtered = new Vector();
	DiligentNode node;
	while (allChildren.hasMoreElements()) {
		node = (DiligentNode) allChildren.nextElement();
		if (!node.isFiltered(true)) filtered.addElement(node);
	}
	return filtered.elements();
}
\end{lstlisting}

\subsection{Overview of the Approach}
\begin{figure}
	\centering
	\fbox{\includegraphics[scale=.6]{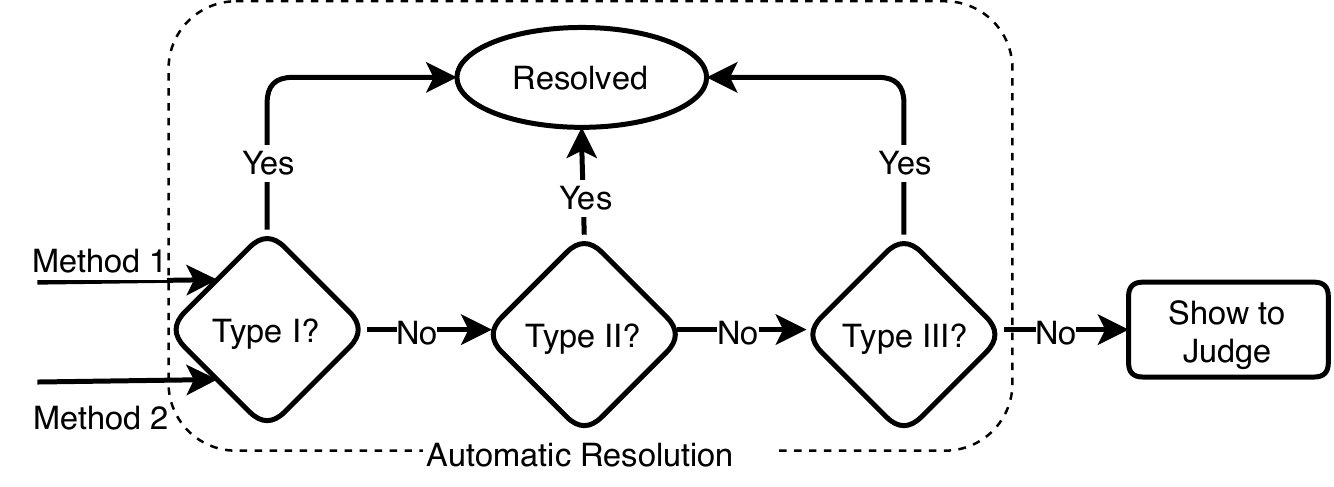}}
	\caption{The pipeline for clone validation.}
	\label{fig:methodology}
\end{figure}

The methodology for clone resolution follows the pipeline presented in
Figure \ref{fig:methodology}. The two methods in a
candidate pair\footnote{A candidate pair consists of two piece of code reported as clone pair by a
	clone detection tool. Our approach validates these pairs, and only when they
	are resolved as true positives they are called clone pairs.} go through a series of steps in which they are checked
against a certain clone type. If at any step a pair is evaluated as a
true clone pair, it is marked as a true positive and the system
proceeds to the next candidate pair. Otherwise, if all of the steps
are failed to evaluate a pair as a true positive, the pair is presented to
the human judge for manual inspection. These steps are ordered by
computational complexity for system performance (and also by
increasing clone type complexity), and are individually described in
the next sections.

\subsection{Automatic Resolution of \tone\ Clones}
\label{subsect:auto_t1}
As described in Section~\ref{sect:defs}, two pairs are \tone\ clones if they are exact replicas
when neglecting source code comments and layout\footnote{The syntax of
  Java is not dependent on layout, so we can ease the definition of
  \tone\ clones. For layout-dependent syntaxes like the ones found in
  Python or Haskell, this approach would require a more careful
  deliberation}. This makes the validation of \tone\ candidates similar to
a simple string comparison after removing certain
elements. We use Algorithm~\ref{alg:autot1} to check if a candidate pair is a \tone\ clone. Starting with
a candidate pair, the algorithm, first, removes all source code comments from both method bodies \textit{(lines 2 and 3)}, then removes white spaces and newlines from them \textit{(lines 4 and 5)}, and finally computes and compares the Hash (\textbf{SHA-256}), of both method bodies \textit{(line 6)}. 

\begin{algorithm}
	\scriptsize
	\textbf{INPUT:} $M1$ and $M2$ are strings representing the method bodies (including method signature) of two methods for which we want to know if they are \tone\ clones. 
	\textbf{OUTPUT:} $Boolean$ \\
	\begin{algorithmic}[1]
		\Function{IsTypeOne}{$M1$, $M2$}
		\LState $M1$ = \Call{RemoveComments}{$M1$}
		\LState $M2$ = \Call{RemoveComments}{$M2$}
		\LState $M1$ = \Call{RemoveWhitespacesAndNewLines}{$M1$}
		\LState $M2$ = \Call{RemoveWhitespacesAndNewLines}{$M2$}
		\LState \Return \Call{Hash}{$M1$}==\Call{Hash}{$M2$}
		\EndFunction 
	\end{algorithmic}
	\caption{\small Automatic \tone\ Resolution}
	\label{alg:autot1}
	
\end{algorithm}
\subsection{Automatic Resolution of \ttwo\ Clones}\label{sec:auto_t2}
To resolve \ttwo\ pairs automatically, we use two heuristics as described below:

\textbf{Action heuristic}: Action tokens of a method form a more stable semantic signature for the method than the identifiers or types chosen by the developer. This is because identifiers and types often change in duplicating methods, while Action tokens tend to remain the same. The reason is that methods and class attributes, represented by Action tokens, bring pre-implemented functionalities, which reduce the burden of coding, and hence, are not probable to be removed or modified after cloning. 

\textbf{Metric heuristic}: Software metrics, measuring different characteristics of source code, can capture structural information of a method. These measurements are resilient to changes in identifier names and literals -- a useful property in the detection of \ttwo\ clones. Hence, we use 24 method level software metrics shown in Table~\ref{tab:jhawk-metrics} for \ttwo\ resolution. The details of these metrics can be found elsewhere~\cite{Saini2018, oreopreprint}. A detailed explanation about the application of Action tokens and software metrics in clone detection can be found in~\cite{oreopreprint}.

\begin{table}[!tbp]
	\begin{center}
		\caption{Method-Level Software Metrics from \cite{oreopreprint}}
		\label{tab:jhawk-metrics}
		\resizebox{\linewidth}{!}{
			\begin{tabular} {l l l l}
				\midrule
				Name & Description &Name & Description \\
				\midrule
				XMET & \# external methods called & HEFF & Halstead effort to implement\\
				VREF & \# variables referenced  & HDIF & Halstead difficulty to implement\\
				VDEC & \# variables declared  & EXCT & \# exceptions thrown\\
				NOS & \# statements  & EXCR & \# exceptions referenced\\
				NOPR & \# operators  & CREF & \# classes referenced\\
				NOA & \# arguments  & COMP & McCabes cyclomatic complexity\\
				NEXP & \# expressions  & CAST & \# class casts\\
				NAND & \# operands  & NBLTRL$*$ & \# Boolean literals\\
				MDN & maximum depth of nesting  & NCLTRL$*$ & \# Character literals\\
				LOOP & \# loops (for,while)  & NSLTRL$*$ & \# String literals\\
				LMET & \# local methods called  & NNLTRL$*$ & \# Numerical literals\\
				HVOC & Halstead vocabulary & NNULLTRL$*$ & \# Null literals\\
				
				
				\thickhline
			\end{tabular}
		}		
	\end{center}
\end{table}

We use Algorithm~\ref{alg:autot2} to check if a candidate pair is a \ttwo\ clone. First,we get a list of action tokens for both methods \textit{(line 2 and 3)}. Then, we compare if these lists are identical \textit{(line 4)}, that is, the contents along with their order of appearance in these lists match. If the lists are identical, we get a list of metrics for both methods \textit{(line 5 and 6)} and then return true if these lists are identical \textit{(line 7)}, else return false. 

This algorithm ensures that a candidate pair is resolved as \ttwo\ only when there is a 100\% match in both the metrics and the Action tokens. The rational is that \ttwo\ clones differ in identifier names and literal values while their structure (captured by metrics), and their method calls and accessed class fields (captured using Action tokens) remain the same. 


\begin{algorithm}
	\scriptsize
	\textbf{INPUT:} $M1$ and $M2$ are strings representing the method bodies (including method signature) of two methods for which we want to know if they are \ttwo\ clones. 
	\textbf{OUTPUT:} $Boolean$ \\
	\begin{algorithmic}[1]
		\Function{IsTypeTwo}{$M1$, $M2$}
		\LState $ListATofM1$ = \Call{GetActionTokens}{$M1$}
		\LState $ListATofM2$ = \Call{GetActionTokens}{$M2$}	
		\If{ \Call{IsIdentical}{$ListATofM1$,$ListATofM2$}}
		\LState $ListMetM1$ = \Call{GetMetrics}{$M1$}
		\LState $ListMetM2$ = \Call{GetMetrics}{$M2$}
		\LState \Return \Call{IsIdentical}{$ListMetM1$,$ListMetM2$}
		\EndIf
		\LState \Return $False$
		\EndFunction 
	\end{algorithmic}
	\caption{\small Automatic \ttwo\ Resolution}
	\label{alg:autot2}
\end{algorithm}

\subsection{Automatic Resolution of~\tthree\ Clones}\label{at_t3}

 
Typically, syntactic clone detectors detect clones in ST3 and VST3 categories. This is because there is a high probability that code snippets with less than 70\% syntactical similarity are coincidentally similar. Moreover, to detect clones between 0-70\% similarity range, detectors may need to capture the semantic similarity, a harder problem in clone detection. Therefore, achieving very high precision in the automatic resolution of all of these subcategories is considered hard. As we want to make sure whatever candidate pair our approach resolves
as a clone pair, is indeed a true clone pair, we focus on the first two subcategories, namely VST3 and
ST3.
 

To resolve~\tthree\ candidates automatically, we
first make the candidate pairs go through an~\AF, and the ones that survive this
filter are fed to a deep learning classifier that predicts
whether they are true clone pairs. Placing~\AF\ before the classifier ensures that candidate pairs, whose methods do not share a specific amount of functionalities, are filtered out early, and shown to judges instead of being resolved automatically; hence, increasing precision.

In the following subsections we first explain the training set used in
training the deep learning model, and next, we describe the details of
the trained model. Finally, we provide the results of a sensitivity
analysis we did for selecting the proper~\AF\ threshold to resolve~\tthree\ clones with high precision.

\subsubsection{Dataset Curation}\label{dataset}

Since the machine learning classifier is supposed to resolve
\tthree\ candidates, we need a training set with clone pairs from
this category. Also, to resolve clone pairs with very high precision, we want our dataset to contain true clone pairs which are very similar in terms of both their semantics and structure. To generate such dataset, we use two token based state of the art clone detectors, CloneWorks (Aggressive mode)
~\cite{svajlenko2017fast} and SourcererCC
~\cite{sajnani2016sourcerercc}. The configurations of these tools are shown in Table~\ref{tab:bcb_manual}. These clone detectors detect \tthree\ clone pairs up to ST3 category, where the methods in each clone pair have high structural similarity. On the other hand, we ensure high semantic similarity in the methods of each clone pair in our dataset by using Action Filter with threshold set to 90\%. The starting dataset used to generate our training dataset is BigCloneBench.

The numbers related to dataset creation process are reported in Table~\ref{tab:dataset}. The training set includes equal number of both assumed positives (clones) and assumed
negatives (non-clones). To get the set of assumed positives, we took an intersection of the clone pairs detected by the two tools (RowId 3 in
Table~\ref{tab:dataset}). We then removed all \tone\ and \ttwo\ pairs from this intersection and selected the pairs which satisfy our Action filter (RowId 4). To ensure that these pairs are true clone pairs, we randomly sampled 1,851 pairs (a statistically significant sample with 99\% confidence
level and 3\% confidence interval), and validated them manually. Two
judges, who are also the authors of this paper, independently went
through these clone pairs and unanimously found all pairs to be true
clone pairs.  


Listing~\ref{lst:vst3} shows an example of true clone pair (VST3) found by the judges. Both methods in this example seem to have a very similar aim: first, they fill a \textit{LinkedList} \textit{(line 6 and 26)} and then they iterate over the \textit{LinkedList} to remove its contents \textit{(lines 14 to 16 and lines 34 to 36)}. The methods not only share many Action tokens, their structures also look very similar. Moreover, the line and token similarity between the methods are high, making this pair a good example of a true positive.

\begin{lstlisting} [label={lst:vst3}, float,floatplacement=H,caption=Example VST3 Clone Pair] 
private boolean doPurgeStudy(DirWriter w, DirRecord parent, int[] counter) throws IOException {
	boolean matchAll = true;
	LinkedList toRemove = new LinkedList();
	for (DirRecord rec = parent.getFirstChild(true); rec != null; rec = rec.getNextSibling(true)) {
		if (doPurgeSeries(w, rec, counter)) {
			toRemove.add(rec);
		} else {
		matchAll = false;
		}
	}
	if (matchAll) {
		return true;
	}
	for (Iterator it = toRemove.iterator(); it.hasNext(); ) {
		counter[0] += w.remove((DirRecord) it.next());
	}
	return false;
}
--------------------------------
private boolean doPurgeInstances(DirWriter w, DirRecord parent, int[] counter) throws IOException {
	boolean matchAll = true;
	LinkedList toRemove = new LinkedList();
	for (DirRecord rec = parent.getFirstChild(true); rec != null; rec = rec.getNextSibling(true)) {
		File file = w.getRefFile(rec.getRefFileIDs());
		if (!file.exists()) {
			toRemove.add(rec);
		} else {
		matchAll = false;
		}
	}
	if (matchAll) {
		return true;
	}
	for (Iterator it = toRemove.iterator(); it.hasNext(); ) {
		counter[0] += w.remove((DirRecord) it.next());
	}
	return false;
}
\end{lstlisting}

\begin{table}
	\centering
	\caption{Dataset Creation Process Statistics}
	\resizebox{\linewidth}{!}{
	\begin{tabular}{clr}
		\toprule
		RowId &	Dataset & Number of Pairs \\
		\midrule
		1	& SourcererCC Pairs & 909,409 \\ 
		2 &	CloneWorks Pairs & 8,053,303  \\
		3 &	SourcererCC \& CloneWorks Intersection & 699,389 \\	
		4 &	Intersection after Removal (Clone Pairs) & \textbf{53,058} \\
		\midrule
		5 &	Non-clone pairs at 90\% \AF & 18,195,489 \\
		6 &	Union of  Pairs by SourcererCC, CloneWorks, Nicad & 8,408,734 \\
		7 &	Non-clones after Removing Union Pairs & 18,135,188 \\
		8 &	Non-clones after Random Sampling & \textbf{53,058} \\ 
		\midrule
		9 &	Total Rows in Final Dataset & \textbf{106,116} \\
		\bottomrule
	\end{tabular}
	}
	\label{tab:dataset}
\end{table}

The training set needs not only positive samples of clones, but also
negative ones. Getting these pairs is considerably more difficult: while there is an
enormous amount of code pairs that are not clones of each other, for
machine learning purposes, it is not useful to include pairs that
have no similarities whatsoever. Ideally, we would like to include
pairs that we know with high certainty are not clones, but that are
sufficiently similar that they could be confused as clones.


To get such assumed negative pairs, we modified Oreo~\cite{oreopreprint}, a clone detector designed to detect \tthree\ clones even in harder clone categories, to predict non-clones such that they have at least 90\%
similarity in their Action tokens (RowId 5). The original source code of Oreo is available at~\cite{saini_vaibhav_2018_1317760}.
Then, we took a union of the clone
pairs reported by three state of the art clone detectors: CloneWorks,
SourcererCC, and NiCad~\cite{roy2008nicad} (RowId 6). NiCad's configurations are shown in Table~\ref{tab:bcb_manual}.
To ensure high confidence in the non-clone pairs, we removed any non-clone pair which is present in the union set (RowId 7). Finally, we did a manual analysis similar to what we did for true positives to gain more assurance about the non-clone pairs. The same two judges, independently as before, went through a random sample of 400 pairs. They found many examples which were definitely non-clones, and also found some examples of MT3 and WT3/4 clones, where the pairs shared high semantic similarity but the structural similarity was weak. This is useful in increasing the precision of the machine learning model since it learns to classify these harder pairs, which are closer to the threshold boundary, as non clones. This is a desirable behavior as these harder cases are then left for human judgment. 

Listing~\ref{lst:copymt3} shows an example of an MT3 pair found by the judges. Both methods in this pair are semantically similar as they both intend to copy the contents from an \textit{InputStream} to an \textit{OutputStream}. The structural and token similarity between the two methods, however, is low, making it harder to detect as a clone pair by many token based clone detectors. Similarly, Listing~\ref{lst:copywt3} shows another pair that semantically, are performing the same task, but the structural similarity between the two methods is very low. Such methods are good candidates that should be left for human judgment.

\begin{lstlisting} [label={lst:copymt3}, float,floatplacement=H,caption=Example MT3 Clone Pair] 
public static void copy(InputStream i, int buf, OutputStream o) throws IOException {
	byte b[] = new byte[buf];
	for (; ; ) {
		int g = i.read(b);
		if (g == -1) break;
		o.write(b, 0, g);
	}
}
--------------------------------
public static void copyTo(InputStream in, OutputStream out) throws IOException {
	byte buffer[] = new byte[2048];
	int n = 0;
	while ((n = in.read(buffer)) != -1) {
		out.write(buffer, 0, n);
	}
	out.flush();
}
\end{lstlisting}

\begin{lstlisting} [label={lst:copywt3}, float,floatplacement=H,caption=Example WT3/4 Clone Pair] 
protected void copy(InputStream _in, OutputStream _out) throws IOException {
	byte[] buf = new byte[1024];
	int len = 0;
	while ((len = _in.read(buf)) > 0) _out.write(buf, 0, len);
}
--------------------------------
public static long copy(InputStream is, OutputStream os) throws IOException {
	byte buffer[] = new byte[1024];
	int readed;
	long length = 0;
	while ((readed = is.read(buffer)) != -1) {
		if (readed > 0) {
			length += readed;
			os.write(buffer, 0, readed);
		} else {
		TimerHelper.notSafeSleep(100);
	}
}
return length;
}
\end{lstlisting}
We then took a random sample of 53,058 pairs (RowId 8) from the above obtained pairs (RowId 7) to have the number of non-clone pairs matched with the number of true clone pairs (RowId 4).
Finally, we aggregated the pairs from RowId 4 and RowId 8 to create a dataset (RowId 9) which we used for training and validating the machine learning model. Each row of this finalized dataset contains a method pair represented as a vector of 48 metrics (24 metrics of Table~\ref{tab:jhawk-metrics} for each method), and a label denoting whether this pair is clone or not.

\subsubsection{Deep Learning Model}\label{ml}

To classify \tthree\ clones we are using Siamese architecture to train a Deep Neural Networks (DNN) model. Siamese models are well suited for problems where two things need to be compared against each other, for example comparing fingerprints~\cite{baldi93finger}. Also, in a recent work on clone detection, Saini et al. compared different architectures of Deep Neural Networks (DNN) and found Siamese DNN to outperform the other DNN architectures~\cite{oreopreprint}. We also carried out model comparison analyses on the train dataset at hand and found Siamese model to outperform other models. Model comparison results are explained later in this section.



\begin{figure}
	\centering
	\fbox{\includegraphics[width=8.6cm,height=4.4cm]{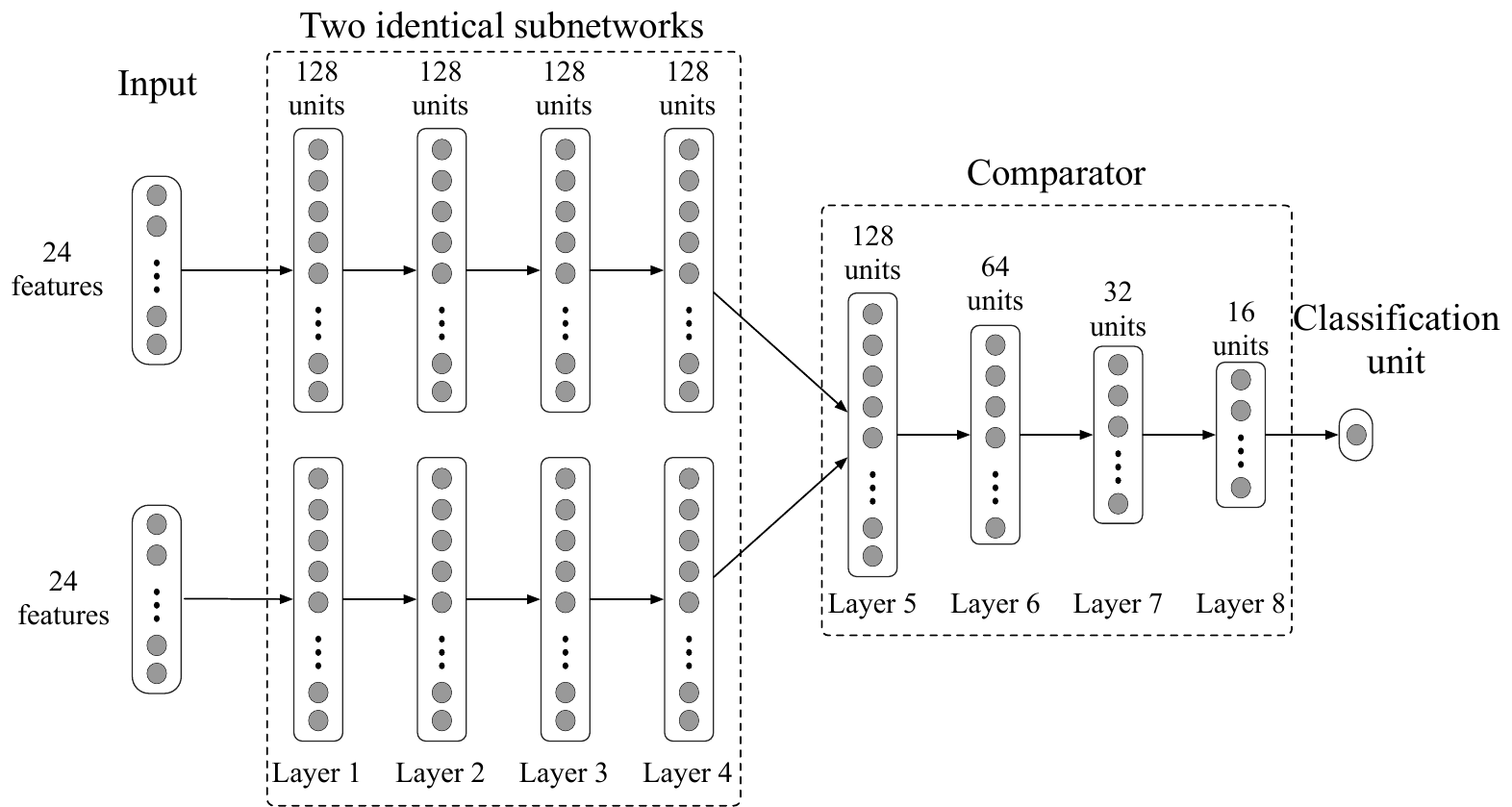}}
	\caption{Siamese architecture}
	\label{fig:Siamese}
\end{figure}

Figure~\ref{fig:Siamese} shows the architecture of the Siamese model
we trained for our approach. It consists of three components: i) two
identical subnetworks, ii) a comparator unit, and iii) a
classification unit. The input to the model are the feature vectors of
each method (24 metrics) in the candidate pair. These feature vectors
are then transformed and processed by the two subnetworks and the
comparator. Finally, the classification unit outputs a number between
0 and 1, representing the probability of a pair being a clone. A more
detailed explanation of this architecture and its components can be
found elsewhere~\cite{oreopreprint}.

\begin{table}
	\centering
	\caption{Precision and Recall on the test set for the Siamese Neural Network model using different thresholds.}
	\label{tab:threshold}
	\begin{tabular}{ccc}
		Threshold &  Precision & Recall \\
		\midrule
		0.6        &     0.970  &  0.920 \\
		0.7        &     0.981   & 0.882 \\
		0.8        &     0.989  &  0.844 \\
		0.9        &     0.996  &  0.700
	\end{tabular}
	\vspace{-0.125in}
\end{table}

To perform the model selection experiments, 
the train dataset 
of 116,000 code pairs 
is randomly divided into 80\% for training and 20\% for
testing. Furthermore, 5,000 pairs from the training set are set aside
for validation purposes, i.e. for hyperparameter tuning. First, we
explored the hyper-parameter tuning for the Siamese model. The best
performing model is the one shown in Figure \ref{fig:Siamese}. Each of
the two subnetwork layers in this model has 4 layers, each with 128
neurons and the comparator has four fully connected layers of sizes
128-64-32-16. The output of the comparator is then fed into a single
classification neuron with sigmoidal (logistic) activation
function. The output of this neuron is a value between 0 and
1. Normally, the values more than 0.5 are assigned label 1, and values
less than 0.5, are assigned label 0. However, since our goal in this
problem is to achieve almost perfect precision (so that the number of
false positives tends to zero), during production (testing), we set
the threshold to announce a pair a true clone pair to be 0.9. The code
pairs which have prediction values between 0 to 0.9 are sent to human
judges for further inspection. Table \ref{tab:threshold} shows the
Precision and Recall values on the test set at different
thresholds. As it is observed, using this deep learning approach
with a threshold of 0.9 yields almost perfect precision (0.996).



\begin{figure*} 
	\begin{minipage}[b]{0.45\linewidth}
		\includegraphics[width=\linewidth,height=5cm]{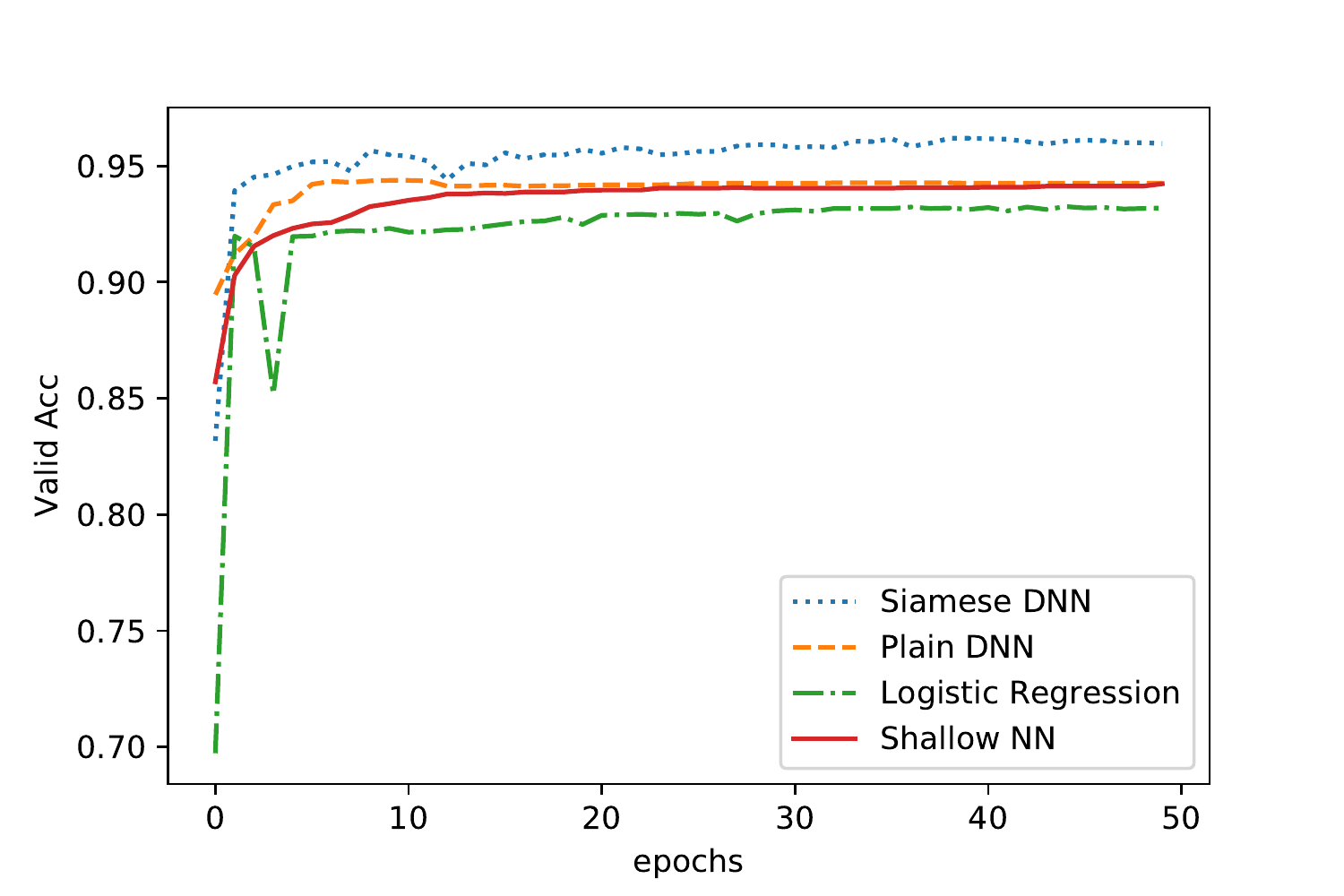} 
		\caption{Validation Accuracy } 
		\label{fig:acc}
	\end{minipage} 
	\hfill
	\begin{minipage}[b]{0.45\linewidth}
		\includegraphics[width=\linewidth,height=5cm]{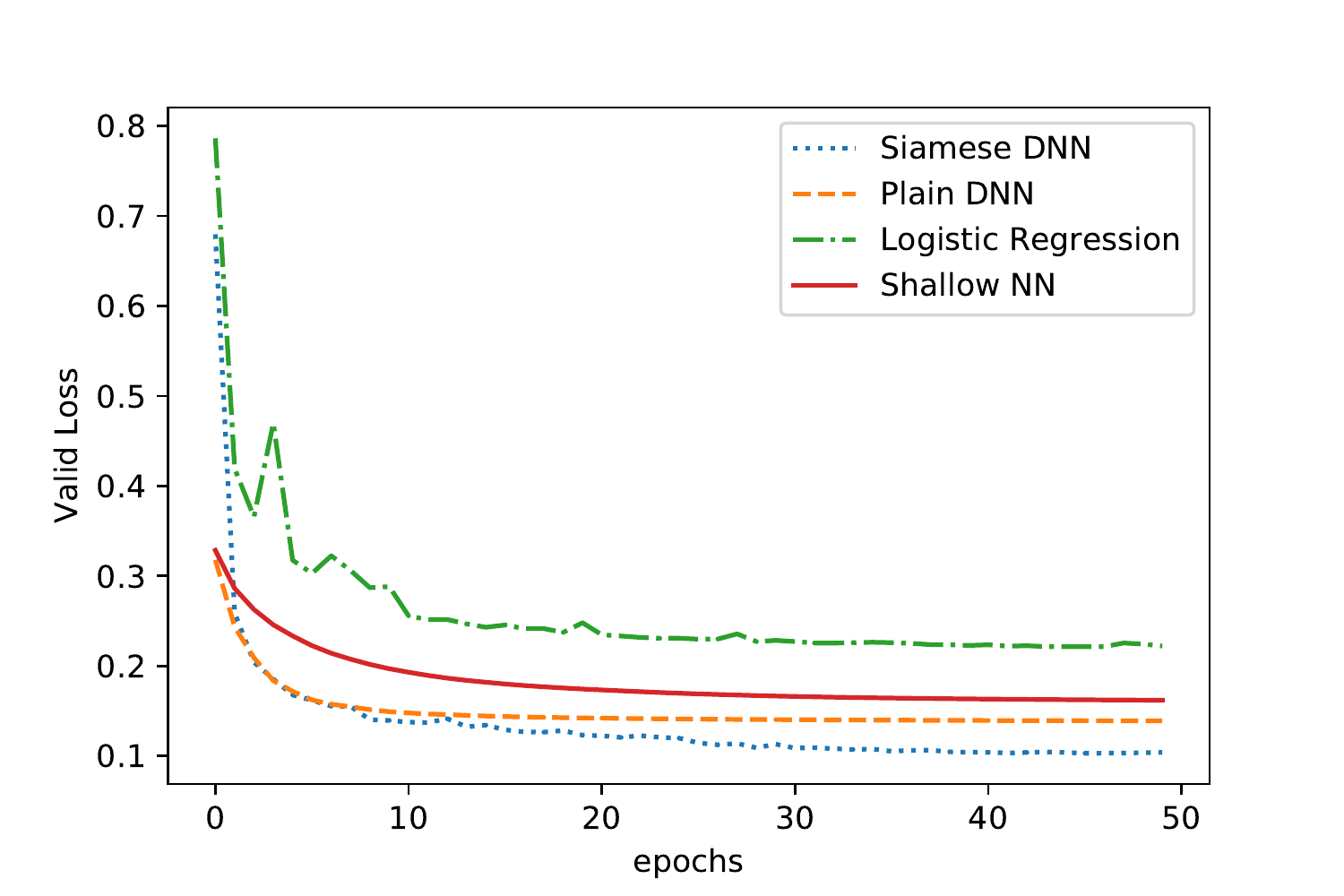} 
		\caption{Validation Loss}  
		\label{fig:loss}
	\end{minipage}
		\vspace{-0.125in}
\end{figure*}

We also compared the Siamese DNN to other models
with different architectures, including: (1) a plain fully connected
neural network (Plain DNN) with similar number and sizes of hidden
layers as the Siamese one; (2) a shallow neural network (Shallow NN)
with one hidden layer and similar number of parameters; and (3) a
logistic regression model. Since our final goal is to achieve a high
Precision, all the comparisons are done when the thresholds for all
models are set to be 0.9. We first compared their performance during
the training stage. Figure \ref{fig:acc} shows that the Siamese DNN
outperforms the other models in terms of accuracy on the validation
set. The accuracy of the Siamese DNN converges to 96.0\%, while the
accuracy for plain DNN and Shallow NN converges only to
94.0\%. Figure \ref{fig:loss} shows that for the validation loss also,
the Siamese structure is superior to the other models. The average
loss value for the Siamese DNN converges to 0.103 (as apposed to Plain
DNN: 0.139; Shallow NN: 0.162; Logistic regression model: 0.222). Thus
in short, the Siamese DNN better fits the training data.

\begin{table}
	\scriptsize
	\centering
	\caption{Precision and Recall values on test set}
	\label{tab:precision}
	\begin{tabular}{lcc}
		Threshold=0.9       & Precision & Recall \\
		\midrule
		Logistic Regression & 0.984     & 0.828  \\
		Shallow NN          & 0.991     & 0.734  \\
		Plain DNN           & 0.983     & 0.841  \\
		Siamese DNN         & 0.996     & 0.700 
	\end{tabular}
	\vspace{-0.125in}
\end{table}

Next, the model performance is compared at the testing
stage. Table \ref{tab:precision} shows that the Siamese network has
the highest Precision, equal to 0.996. In short, this shows that the
Siamese network has better generalization performance than the other
models used in the comparison.



\subsubsection{Sensitivity Analysis}\label{sens_analysis}

We did a sensitivity analysis to find the optimum threshold of \AF\ with the goal of not having any false positives, and maximizing the number of pairs resolved automatically.
 
\textbf{Methodology}. We used the clone pairs reported by SourcererCC on the BigCloneBench Dataset. We ran InspectorClone with four different threshold values of \AF: 60\%, 65\%, 70\%, and 75\%. One author manually inspected the clone pairs that are automatically resolved by \IC\ to figure out the number of false positives in them.
Results of this analysis are denoted in Table
~\ref{tab:sens}. The first column of this table shows the examined
thresholds, and the next three columns, respectively, denote the
number of automatically resolved \tone, \ttwo, and \tthree\ clone pairs. The next three columns show
the number of false positives observed at each clone category, and the
last column depicts the number of clone pairs that need the manual
validation by humans. At 60\% and 65\% thresholds, we observed some false positives, whereas at 70\% and 75\% thresholds, no false positives were observed. The number of automatically resolved clone pairs at 70\% threshold (49) is greater than this number at 75\% threshold (18). Consequently, 70\% threshold was selected to be used in \AF.



\begin{table}
	\centering
	\caption{Sensitivity Analysis Statistics}
	\begin{tabular}{cccccc}
		\toprule
		Threshold& Auto & Auto   & Manual \\
		& \tthree\ & \tthree\ FP & \\
		\midrule
		60\% & 124 & 16  & 68 \\
		65\% & 78 & 17  & 121 \\
		70\% & 49 &  0  &149 \\
		75\% & 18 &  0  & 174 \\
		\bottomrule
	\end{tabular}
	\label{tab:sens}
	\vspace{-0.15in}
\end{table}

\section{Evaluation}
\label{sect:eval}

As discussed earlier, the main goal of our approach is to automatically resolve as much as clone pairs as possible with high precision. Hence,
to evaluate it, we designed an experiment using
\IC, and seven clone detectors.  The goal of
this experiment is twofold: i) to understand the impact of our
approach on the reduction of manual effort, and ii) to measure
the precision of the automatic clone resolution approach.

We include SourcererCC, iClones~\cite{gode2009incremental}, NiCad, CloneWorks, SimCad~\cite{uddin2013simcad} as popular examples of modern clone detectors that support \tthree\ clone detection. CloneWorks comes in two different modes, Aggressive and Conservative; we tested \IC\ on both of these modes. We also include two recent tools, Oreo and CCAligner~\cite{Wang:2018:CTB:3180155.3180179}. While all these tools detect clones in \tone, \ttwo, and early categories of \tthree\ (\textit{VST3} and \textit{ST3}), Oreo and CCAligner are capable of detecting clones beyond \textit{ST3} categories such as \textit{MT3}. We also wanted to include Deckard~\cite{jiang2007deckard} and CPD~\cite{cpd}; however, they both detect clones beyond method boundaries. At this time, we cannot reliably conduct a meaningful experiment with them on \IC, which only supports method level clone detectors as of now. Also, both of these tools report their results as clone classes and not as clone pairs. When we ran processes to generate clone pairs from these clone classes, they both produced large amount of clone pairs. We killed the processes after generating more than 175G of clone pairs for each of them as these are very big files for \IC\ to process. 

We ran all tools on the recall dataset of BigCloneBench and obtained the clone pairs reported by each tool. We then uploaded the clones reported by each tool to \IC, and calculated the number of clone pairs automatically resolved in each category, and the number of pairs left for manual validation. We configured \IC\ to consider only those pairs that have methods with at least 50 language tokens, a standard size filter used in precision studies~\cite{sajnani2016sourcerercc,oreopreprint}. 

To gain high confidence in our experiment results,
we conducted 2 rounds of experiments for each tool (a total of
16 experiments, with 7 tools and CloneWorks being executed in two modes). In each round, \IC\ sampled 400 random candidate pairs
from the output of each tool. \IC\ then automatically resolved
some clone pairs as assumed positives, leaving the rest for manual validation. To measure the precision
of the automatic resolution part, five judges, who are
also authors of this paper, independently went through the whole set
of 2,545 automatically resolved clone pairs to look for possible false
positives. The judges were also asked to report the time they took to complete each round of experiment. In total, it took around $58$ person hours to complete
all 80 experiments (16 rounds per each judge).

\begin{table}[t]
	\scriptsize
	\centering
	\tabcolsep=0.075cm
	\centering
	\caption{Reduction of Manual Effort}
	\begin{tabular}{lcccccccccl}
		\toprule
		\multicolumn{1}{l}{Tool} && \multicolumn{3}{c}{Automatically} && \multicolumn{1}{c}{Manual} && \multicolumn{1}{c}{FP} &&\multicolumn{1}{c}{Tool Configuration} \\
		&& \multicolumn{3}{c}{Resolved} && Inspection && && \\
		\cmidrule{3-5} 
		&&\multicolumn{1}{c}{T1}  & \multicolumn{1}{c}{T2} & \multicolumn{1}{c}{T3} && (out of 400) && && \\
		\cmidrule{1-1} \cmidrule{3-5} \cmidrule{7-7} \cmidrule{9-9} \cmidrule{11-11}
		\multirow{2}{*}{CCAligner} &&18 & 38 & 60 && 284 &&0 &&MIL=6, $\Theta=60\%$\\
								   &&18 & 31 & 64 && 287 &&0 &&e=1, q=6 \\
		\cmidrule{1-1} \cmidrule{3-5} \cmidrule{7-7} \cmidrule{9-9}
		\cmidrule{11-11}
		\multirow{2}{*}{CloneWorks(A)} &&118 & 27 & 34  && 221&&0 && MIT=1, $\Theta=70\%$,\\
									   &&118 & 35 & 37  && 210&&0 &&Mode=Aggressive \\
		\cmidrule{1-1} \cmidrule{3-5} \cmidrule{7-7} \cmidrule{9-9}	
		\cmidrule{11-11}
		\multirow{2}{*}{CloneWorks(C)} &&53 & 43 & 36  && 268&&0&& MIT=1, $\Theta=70\%$,\\
									   &&54 & 58 & 26  && 262&&0&&Mode=Conservative \\
		\cmidrule{1-1} \cmidrule{3-5} \cmidrule{7-7} \cmidrule{9-9}
		\cmidrule{11-11}
		\multirow{2}{*}{iClones} && 254 & 59 & 26 && 61 &&0&&MIT=50,\\
		                         && 256 & 63 & 15 && 66 &&0&&min block=20\\
		\cmidrule{1-1} \cmidrule{3-5} \cmidrule{7-7} \cmidrule{9-9}
		\cmidrule{11-11}
		\multirow{2}{*}{NiCad} && 99 & 35 & 159 && 107 &&1&&MIL=6, BIN=True,\\
		                       && 115 & 26 & 165 && 94 &&0&&IA=True, $\Theta=30\%$\\
		\cmidrule{1-1} \cmidrule{3-5} \cmidrule{7-7} \cmidrule{9-9}
		\cmidrule{11-11}
		\multirow{2}{*}{Oreo} && 0 & 0 & 0 && 400 &&0&&MIT=15,$\Theta=55\%$,\\
		                      && 0 & 0 & 0 && 400 &&0&&$\Gamma=60\%$ \\
		\cmidrule{1-1} \cmidrule{3-5} \cmidrule{7-7} \cmidrule{9-9}
		\cmidrule{11-11}
		\multirow{2}{*}{SourcererCC} && 155 & 25 & 8 && 212&&0 &&MIT=1,\\
									 && 149 & 24 & 12 && 215&&0 &&$\Theta=70\%$\\
		\cmidrule{1-1} \cmidrule{3-5} \cmidrule{7-7} \cmidrule{9-9}
		\cmidrule{11-11}
		\multirow{2}{*}{SimCad} && 15 & 0 & 2 && 383 && 0&&GT=True,\\
								&& 10 & 2 & 3 && 385 && 0&&US=True,MIL=6 \\
		\bottomrule
	\end{tabular}
		\vspace{-0.2in}
	\label{tab:bcb_manual}
\end{table}





The results of this experiment are shown in
Table~\ref{tab:bcb_manual}. The first column shows the name of the tool. The next three columns denote, respectively, the
number of \tone, \ttwo, and \tthree\ candidate pairs that were
automatically resolved by \IC. The fifth column shows the number of candidate pairs that could not be automatically classified, and needed manual validation by humans (out of the sample of 400). The sixth column (\textit{FP}) contains the number of false positives (after considering majority vote) observed by human judges in the automatically resolved pairs. And finally, the seventh column shows the configurations which were used to run the tools. These configurations are based on our discussions with their developers, and also the configurations suggested in~\cite{Svajlenko:2015:ECD:2881297.2881379}. In the table, \textit{MIT} stands for minimum tokens, \textit{MIL} stands for minimum number of lines, \textit{BIN} and  \textit{IA},  respectively stand for blind
identifier normalization and literal abstraction used in NiCad. $\Theta$ stands for similarity threshold (for NiCad, it is \textit{difference threshold}, and for Oreo it is \AF\ threshold),
$\Gamma$ is the threshold for input partition used in
Oreo. In CCAligner's configurations, \textit{e}  stands for \textit{edit distance}, and \textit{q} is the \textit{window size}. \textit{GT} and and \textit{US} stand for \textit{greedy transformation} and \textit{unicode support} used in SimCad.

As the table shows, \IC\ reduced the number of pairs that
need manual analysis for all tools except for Oreo. On an average, there is a 39\% reduction in the number of clone pairs that are left for human judges. Most reduction is observed for iClones (84\%) and NiCad (74\%), while for SimCad (4\%) and Oreo (0\%), we observed little to no reductions. The reduction for rest of the tools ranges from 28\% to 47\%.

To understand why \IC\ did not help in reducing the number of pairs for Oreo and SimCad, two judges went through the samples of one of the two experiments conducted for both tools. For SimCad, the judges reported 358 out of 400 pairs as false positives of the SimCad tool itself (10.5\% precision). The presence of large number of false positives in the pairs of SimCad  explains why \IC\ did not help much in resolving its pairs.
\begin{lstlisting} [label={lst:oreo}, float,floatplacement=H,caption=Example Candidate Pair from Oreo] 
public static <T> T readStreamAsObject(InputStream inputStream, Class<T> type) throws ClassNotFoundException, IOException {
	  ObjectInputStream objectInputStream = null;
	  try {
		  objectInputStream = new ObjectInputStream(inputStream);
		  return type.cast(objectInputStream.readObject());
	  	} finally {
	  Utility.close(objectInputStream);
   }
}
--------------------------------
public static <T extends Serializable> T deserialise(Class<T> class1, File out) throws ClassNotFoundException {
	  try {
	  	  FileInputStream fis = new FileInputStream(out);
		  ObjectInputStream in = new ObjectInputStream(fis);
		  Object output = in.readObject();
		  in.close();
		  return class1.cast(output);
	  	} catch (IOException ex) {
	  ex.printStackTrace();
	  return null;
   }
}
\end{lstlisting}
For Oreo, the judges reported a much higher precision of 80\%, where they reported 80 out of 400 pairs as false positives of the tool. Almost all of the clone pairs in the sample of Oreo were found to be in harder to detect \tthree\ categories (\textit{MT3} and \textit{WT3}). An example of such a pair is shown in Listing~\ref{lst:oreo}. Both methods in this example are reading an \textit{object} from an input stream, and then, they cast this object into the \textit{type} they received in their arguments. Though they are performing similar tasks and hence, are semantically similar, they differ significantly in their structural properties. This qualifies such pairs to fall in harder to detect MT3/WT3 categories, making them good candidates for human inspection.

If we remove Oreo and SimCad, which are two special cases, from the analysis, on an average, \IC\ resolves 52\% of the clone pairs. The results demonstrate that \IC\ can have a key role in reducing the burden of manual effort needed by users in precision studies. 

Apart from the reduction in manual effort, the precision of the
automatic classification is of a great importance. Out of 1,432 \tone, and 466 \ttwo\ clone pairs resolved by \IC, judges found no false positives, giving \IC\ perfect precision scores in these categories. The judges reported some false positives in the \tthree\ pairs. We report the precision for \IC\ with following two strategies: i) Strategy-A, when majority vote is considered (numbers in column 6 of Table~\ref{tab:bcb_manual} are based on this strategy), and ii) Strategy-B, when a pair is considered false positive if any of the judges report it as a false positive. In Strategy-A, one false positive was found out of 647 \tthree\ pairs, giving \IC\ a precision score of 99.8\%. With this strategy, the precision of \IC\ for all types of pairs combined (2,545 pairs) is 99.96\%. The methods in this false positive pair are big in size (	$\approx$140 \textit{NOS}). Both of these methods make around 100 calls to \textit{add()} method of a list object, which results into a high match in their Action tokens. Also, the arguments to these \textit{add()} method calls in both of these methods are \textit{String Literals}, thereby increasing the match count in the \textit{NSLTRL} metric, which in turn contributes to a high structural match, making \IC\ resolve the pair as a \tthree\ clone. However, the \textit{String Literals} are very different and there exists a \textit{loop} in one of the methods, which led the judges to mark this pair as a false positive.

In Strategy-B, 11 false positives were found, giving \IC\ a precision score of 98.3\% in \tthree\ pairs. If pairs of all types are combined, this strategy gives a precision score of 99.57\%. In their judgments of 2,545 pairs, the judges were unanimously in agreement on 2,534 pairs, giving a conservative estimate of inter-rater-reliability as 99.57\%.

When asked about these false positives, all judges mentioned that except for two or three pairs, all of these pairs are borderline cases. For instance, one judge noted: \textit{"I am on the fence about this pair"}. And for a different pair another judge noted: \textit{"I hesitate if it is a clone or not"}. This shows that identification of clones is a subjective task which involves cases that are hard to judge even by humans. We note, that the judges are well aware of the clone definition and clone types and all of them have previously contributed to the research involving software clones or clone detectors. 

The results show that the strict thresholds used for automatic clone validation are appropriate, if not prefect, and that we can rely on the automatically resolved pairs with high confidence.

\section{Related Work} \label{sect:related}
Measuring the detection capabilities of clone detection tools is an important part of source code cloning research. This demands the existence of labeled and standardized datasets that can assist with this measurement. Therefore, development of such datasets have been the focus of research throughout the years. Unlike the vast majority of areas for which the tasks for producing labeled datasets
are accessible to a large number of people (without any
special expertise being required), developing labeled datasets related to source code cloning requires significant expertise in a narrow
topic: programming. For example, image or speech
recognition can be done by everyone; some examples of platforms that assist people with these tasks are Amazon Mechanical Turk, or the population of College
students. However, such platforms cannot be used in preparation of source code cloning datasets due to the need for the related knowledge. For this reason, researchers have tried to build
such labeled datasets in other ways. Most of these works have been successful in estimating the recall since recall estimation does not require the comprehensive labeling of all pairs, which is needed in measuring the precision. Here, we briefly discuss a set of these efforts.

BigCloneBench (BCB) dataset~\cite{Svajlenko:2014:TBD:2705615.2706134, Svajlenko:2015:ECD:2881297.2881379} is probably the most related work to ours. We have also used it in the evaluation of our approach with \IC. 
The underlying corpus of Java source code used by BCB is
IJaDataset-2.0\footnote{Available at
	\url{https://sites.google.com/site/asegsecold//projects/seclone}
	(March 2018).}. This dataset represents a large inter-project Java
repository containing 25,000 open source projects, with 2.3 million
source files and 365M lines of code~\cite{Svajlenko:2015:ECD:2881297.2881379}. 
BCB contains a subset of IJaDataset curated by human judgment, and it
contains over 8 million known clone pairs within IJaDataset. It is
the result of using IJaDataset, selecting a series of known algorithms (sorting algorithms is one example), and tracking possible implementations of these across the dataset. Hence, not all possible clone pairs are tagged in this dataset, and there exists many pairs that are not tagged. As a result, this dataset cannot be used to measure the precision of clone detectors, but it has been used by BigCloneEval (BCE)~\cite{7816515} to estimate the recall of clone
detector tools automatically.
	
	
	

Another dataset is created by Bellon \textit{et al.}~\cite{bellon2007comparison}. To prepare this dataset, Bellon manually validated 2\% of the clones reported by then (year 2002) contemporary clone detectors for eight software systems.
Svajlenko \textit{et al.}~\cite{Svajlenko:2015:ECD:2881297.2881379} found that this benchmark is not suitable for accurate evaluation of modern clone detection tools. They attributed many of the problems in the dataset to it being built using tools that are now outdated. It 
has also been found to have
other problems as we see next.

Murakami \textit{et al.}'s dataset~\cite{murakami2014dataset} is an improvement on the Bellon \textit{et al.}'s dataset. Murakami \textit{et al.} found out that since Bellon dataset does not contain locational information of gaped lines (i.e. lines that are present in a pair but missing in the other), it has not evaluated some \tthree\ clones correctly. Hence, they added this information and improved the dataset with this respect. 

Another effort has been made in SOCO 2014~\cite{soco}. SOCO was a challenge defined for detection of source code pairs that are reused. The task was carried out at document level, and in C/C++ and Java. Two datasets were provided: train and test. Train dataset was labeled and used to train an algorithm that can find source code pairs in which one pair is developed reusing the other one. The test dataset was used to evaluate the accuracy of the developed algorithm with respect to recall, precision, and F1\footnote{The F1 score is a measurement provided by the harmonic mean between precision and recall.}. 
SOCO contains only 259 Java files and 79 C files, and these examples do not represent realistic software projects (the origin of the source code is unclear).

\section{Threats to Validity and Limitations}
\label{sect:threat}

The measurement accuracy of our approach, and its reduction in manual
effort was performed manually and independently by five expert judges
over a large sample of clone pairs detected by seven different clone detection tools. However, these five judges were also
authors of this work and more importantly, like any work that relies
on human action, practical limitations related to bias and cognition
could have affected our analysis. We mitigated this issue by strictly
adhering to the definition of the clone types during manual classification and also by sharing the data for researchers to verify.

The tools used to generate clone pairs and validate our approach can have 
an impact on the validation of our approach. For example, if a tool has a tendency to detect large clones, then the validation will be performed on the large clones too. To compensate for this bias and to gain more confident in our approach we evaluated it with seven different clone detectors.

Another important consideration is that our approach focuses on Java
methods and is evaluated for methods with 50 tokens on more. It is
possible to apply this methodology to other granularities of source
code and to methods smaller than 50 tokens, but doing so would require
modifying the existing components of our approach, specifically the
software metrics.

We measure manual effort involved in precision studies as the number of clone pairs that need manual inspection. The effort, however, to inspect clone pairs of different types and sizes varies significantly and therefore may not be linear to the number of pairs. 

\section{Conclusions and Future Work}
\label{sect:conclusion}

We have presented a semiautomated approach and a tool, \IC, that facilitate
 precision studies. We evaluated the precision of the automatic clone resolution part of this approach on seven different clone detectors. Our experiments show that the precision of \IC\ is very high (>99.5\%) making it suitable for conducting  precision studies. Further, we demonstrated that the number of clone pairs resolved by \IC\ is significant. 
 
 \IC\ is available to the community and it provides a beneficial framework to access community efforts and to contribute back to them.

As future work, we are looking at the implementation of this approach
in different programming languages, different granularities (classes
or files instead of methods for example) and different scales (clone
with less than 50 tokens). 

\bibliographystyle{ieeetr}
\bibliography{bib/bib}

\end{document}